# Pembobolan website KPU (Komisi Pemilihan Umum) Apakah melanggar UU RI no.36 tahun 1999 tentang telekomunikasi ?

Oleh : Spits Warnars Harco Leslie Hendric, S.Kom
- Dosen Tetap Fakultas Teknologi Informasi, Universitas Budi Luhur
- Mahasiswa Magister Teknologi Informasi, Universitas Indonesia

## Abstract

Information Technology KPU (Indonesia Electoral Commision) is a project in supporting democratization process in Indonesia. It is a part of General Election program of KPU-Indonesian Government. The aim of IT KPU is to build the transparency of the ballot result to the public (citizen and international world) and as the embrio of e government in Indonesia. It also has the aim for influence the citizen with Information Technology and the use of computer.

## Pendahuluan

Dani seorang mahasiswa Fakultas Ilmu Sosial dan Politik Universitas Muhammadiyah Yogyakarta (UMY) jurusan Hubungan Internasional yang berasal dari Kebumen, Jawa Tengah mengaku merasa tertantang dengan pernyataan tim TI (Teknologi Informasi) KPU yang bernilai 152 miliar rupiah, yang dengan gamblangnya menyatakan bahwa sistem keamanan KPU sangat kuat dan tak mungkin kena hack. Akhirnya situs penghitungan hasil pemilu di http://tnp.kpu.go.id bobol dan berhasil dihack pada tanggal 17 April 2004 dan tampilan 24 parpol peserta pemilu diubah
Dani yang bekerja sebagai konsultan TI PT Danareksa ini dijerat dengan pasal 22,38 dan 50 UU no.36 Tahun 1999 tentang Telekomunikasi.

### Kronologi Pembobolan

Serangan terhadap TI KPU itu dilakukannya sebanyak dua kali.

1. 16 April 2004 sekitar pukul 01.43 WIB.
    - tes terhadap sistem keamanan kpu.go.id melalui cross site scripting.
    - menggunakan internet protocol (IP) public PT Danareksa.
    - serangan pertama itu gagal.
    - menggunakan IP milik Warna Warnet yang berada di Jl Kaliurang km 8, Yogyakarta.

- menggunakan nama XNUXER.
2. 17 April 2004 pukul 03.12 WIB
    - menyerang lagi server tnp.kpu.go.id dengan cara SQL Injection (menyerang dengan cara memberi perintah melalui program SQL). Contoh: http://tnp.kpu.go.id/DPRDII/dpr_dapil.asp?type=view&kodeprop=1&kodeprop=1&kodekab=7;UPDATE partai set nama='partai dibenerin dulu webnya' where pkid=13;. Penambahan kode SQL tersebut telah menyebabkan perubahan pada salah satu nama partai di situs TNP KPU menjadi 'partai dibenerin dulu webnya'

        Terdakwa berhasil melakukan perubahan pada seluruh nama partai di situs TNP KPU pada jam 11:24:16 sampai dengan 11:34:27. Perubahan ini menyebabkan nama partai yang tampil pada situs yang diakses oleh publik, seusai Pemilu Legislatif lalu, berubah menjadi nama-nama lucu seperti Partai Jambu, Partai Kelereng, Partai Cucak Rowo, Partai Si Yoyo, Partai Mbah Jambon, Partai Kolor Ijo, dan lain sebagainya.

    - Berhasil menembus IP TNP KPU 203.130.201.134.
    - menggunakan teknik spoofing atau penyesatan. Caranya, Dani melakukan hacking dari IP 202.158.10.117 PT Danareksa, kemudian ia membuka IP 208. 147.1.1 Proxy Anonymous Thailand, yang didapatkan dari http://www.samair.ru/proxy
    - Melalui IP itulah, Dani masuk ke IP tnp.kpu.go.id dan berhasil mengubah tampilan nama 24 partai.
    - Awalnya, ia bermaksud mengubah hasil perolehan suara dengan cara jumlah perolehan suara dikalikan 10. Namun gagal, karena field jumlah suara tidak sama dengan field yang tersangka tulis dalam sintaks penulisan.

*Kronologi Penangkapan Si pembobol*

KPU melapor polisi dan segera polisi pun segera bertindak. Enam anggota satuan Cyber Crime Polda Metro Jaya berangkat ke Yogyakarta guna mengecek alamat IP yang didapat dari data KPU. Menurut data itu, pelaku melakukan kegiatan dari IP Address 202.158.10.117 dan berupaya menambah tulisan.

Penyidik melacak pemilik blok IP Address melalui arin.net dan situs www.apnic.net/apnicbin/whois. Sedangkan, untuk melacak rute IP Address, penyidik mencarinya di www.level3.comGlass dan www.apjii.or.id/tools/lg.php Di Yogyakarta, Dani tinggal di Jl Pamularsih No 8 Patang Puluhan Wirobrajan. Berdasarkan catatan polisi, Dani adalah hacker yang berasal dari Yogyakarta dan pindah ke Jakarta sejak 1 April 2003.

Polisi mengamankan barang bukti *router, log file kabinet, server* warnet Yogyakarta, *server* Danareksa, *server* KPU, grafik koneksi berupa *webalizer, cd sofware, boks fileM* dan buku komputer.

## **Permasalahan**
Apa yang dilakukan oleh Dani apakah melanggar UU RI no.36 tahun 1999 tentang telekomunikasi ?

## **Pembahasan**
Menurut saya Dani ***tidak melanggar UU RI no.36 tahun 1999***
Sesuai dengan tulisan diatas Dani akan dituntut dengan pasal 22,38 dan 50 UU RI No.36 tahun 1999. Secara jelas dan gambling kalimat berikut akan menjelaskan per pasal yang dituduhkan kepada Dani.

Pasal 22 berbunyi:
Setiap orang dilarang melakukan perbuatan tanpa hak, tidak sah atau memanipulasi:
    a. akses ke jaringan telekomunikasi, dan atau
    b. akses ke jasa telekomunikasi; dan atau
    c. akses ke jaringan telekomunikasi khusus

Isi pasal akan bisa menjatuhkan jika diartikan secara per kalimat contoh Kalimat "Setiap orang dilarang melakukan perbuatan tanpa hak, tidak sah atau memanipulasi". Jika hanya penggalan kalimat ini yang dijadikan sebagai acuan untuk menjerat Dani jelas-jelas isi pasal ini terkesan dipermainkan !! Dan Jelas kelihatan bahwa hukum hanya berpihak kepada siapa yang mempunyai uang dan kekuasaan.

Jika isi pasal dibaca secara keseluruhan dan diartikan secara keseluruhan mungkin akan menjadi pertanyaan-pertanyaan dan bila melihat kalimat "Setiap orang dilarang melakukan perbuatan tanpa hak, tidak sah atau memanipulasi". Berarti ada 3 kalimat mendasar yang bisa dijadikan dasar pertanyaan yaitu :
- perbuatan tanpa hak
- tidak sah
- manipulasi

Untuk selanjutnya kalimat tidak sah diartikan sebagai "perbuatan tidak sah" dan kata manipulasi diartikan sebagai "perbuatan manipulasi".
ketiga 3 kata mendasar tersebut dipasangkan dengan 3 kalimat yaitu :
- akses ke jaringan telekomunikasi
- akses ke jasa telekomunikasi
- akses ke jaringan telekomunikasi khusus

Rumus perkalian pada matematika 3 kali 3 adalah 9, jadi ada 9 pertanyaan mendasar yang perlu diperjelas yaitu :
Apa yang disebut dan dikategorikan sebagai :
    1. Kalimat Perbuatan tanpa hak
- Perbuatan tanpa hak akses ke jaringan telekomunikasi
- Perbuatan tanpa hak akses ke jasa telekomunikasi

- Perbuatan tanpa hak akses ke jaringan telekomunikasi khusus
2. Kalimat perbuatan tidak sah
    - Perbuatan tidak sah akses ke jaringan telekomunikasi
    - Perbuatan tidak sah akses ke jasa telekomunikasi
    - Perbuatan tidak sah akses ke jaringan telekomunikasi khusus
3. Perbuatan manipulasi
    - Perbuatan manipulasi akses ke jaringan telekomunikasi
    - Perbuatan manipulasi akses ke jasa telekomunikasi
    - Perbuatan manipulasi akses ke jaringan telekomunikasi khusus

Dari 9 kalimat pertanyaan diatas, timbul sebuah pertanyaan, apa yang dimaksud
- Jaringan telekomunikasi
- Jasa telekomunikasi
- Jaringan telekomunikasi khusus

Pada UU RI No. 36 Tahun 1999 yang dimaksud dengan
1. Jaringan Telekomunikasi
   Pasal 1 UU RI NO. 36 tahun 1999 menjelaskan tentang arti kata kalimat jaringan telekomunikasi sebagai berikut :
   "Jaringan telekomunikasi adalah rangkaian perangkat telekomunikasi dan kelengkapannya yang digunakan dalam bertelekomunikasi"
2. Jasa Telekomunikasi
   Pasal 1 UU RI NO. 36 tahun 1999 menjelaskan tentang arti kata kalimat jasa telekomunikasi sebagai berikut :
   "Jasa telekomunikasi adalah layanan telekomunikasi untuk memenuhi kebutuhan bertelekomunikasi dengan menggunakan jaringan telekomunikasi"
3. Jaringan Telekomunikasi Khusus
   Pada UU RI No.36 tahun 1999 ini tidak ada kalimat yang menjelaskan tentang jaringan telekomunikasi khusus ini, namun bila melihat penjelasan tentang arti kalimat Jaringan telekomunikasi, maka bila ditambahkan dengan kata khusus, berarti ada sesuatu jaringan telekomunikasi yang dikhususkan. Kata arti khusus ini bisa lebih menjurus kepada Jaringan Telekomunikasi Negara.

Setelah kita mengetahui arti kata dari ke-3 kalimat diatas, berikutnya kita mengambil kata "akses" mencari apa yang dimaksud dengan kata akses !. Pada UU RI No.36 tahun 1999 ini tidak ada kalimat atau pasal yang menjelaskan tentang pengertian atau istilah apa yang dimaksud dengan kata "akses".

Saya belum mendapatkan istilah yang tepat untuk kata "akses", namun jika dihubungkan dengan dunia internet akan ada istilah-istilah berikut yang cukup membantu kita apa yang dimaksud dengan kata akses :
- Jika anda ingin mengerjakan tugas e-assigment silahkan akses ke website saya di [www.spits.8k.com](www.spits.8k.com) !!
- Jika ingin cari berita yang up todate akses ke [www.detik.com](www.detik.com)
- Dan seterusnya !!

Dari kalimat-kalimat diatas yang mengandung kata "akses" dan dihubungkan dengan dunia internet, maka kita berkesimpulan bahwa kata akses mempunyai arti kata :
"sebuah kegiatan untuk dapat menampilkan sebuah halaman web dengan menggunakan software browser"

Dari semua istilah-istilah yang sudah kita artikan, maka marilah kita coba menjawab ke-9 pertanyaan, diantaranya:
- Perbuatan tanpa hak akses ke jaringan telekomunikasi
- Perbuatan tanpa hak akses ke jaringan telekomunikasi khusus
- Perbuatan tidak sah akses ke jaringan telekomunikasi
- Perbuatan tidak sah akses ke jaringan telekomunikasi khusus
- Perbuatan manipulasi akses ke jaringan telekomunikasi
- Perbuatan manipulasi akses ke jaringan telekomunikasi khusus

Mengacu pada arti kalimat "jaringan telekomunikasi" pada Pasal 1 UU RI NO. 36 tahun 1999 yang menjelaskan tentang arti kata kalimat jaringan telekomunikasi sebagai berikut :
"Jaringan telekomunikasi adalah rangkaian perangkat telekomunikasi dan kelengkapannya yang digunakan dalam bertelekomunikasi"
Kalau bicara Jaringan atau dalam bahasa inggrisnya Networking atau Net atau yang dulu pernah diistilahkan dengan jaring laba-laba. Jika melihat penjelasan pada istilah "rangkaian perangkat telekomunikasi dan kelengkapannya" berarti penekanannya pada hubungan atau rangkaian dari beberapa perangkat telekomunikasi, dimana rangkaian telekomunikasi ini dirangkaikan atau dihubungkan dengan jalur pengiriman data baik melalui kabel atau Wireless.
      Jadi perbuatan tanpa hak akses atau tidak sah atau manipulasi yang bagaimana yang dikategorikan sebagai perbuatan tanpa hak akses/tidak sah/manipulasi ke jaringan KPU. Kita tahu setiap orang yang terhubung dengan internet berhak untuk mengakses http://tnp.kpu.go.id.
Lalu apakah khusus untuk Dani tidak punya hak akses ke http://tnp.kpu.go.id ? Menurut saya apa yang dilakukan oleh Dani dengan men-deface KPU adalah mengakses komputer dimana data-data tentang hasil Pemilu terekam, dan pada UU RI No.36 tahun 1999 tidak ada kalimat atau pasal yang menjelaskan tentang kalimat sebagai perbuatan tanpa hak akses/tidak sah/manipulasi ke komputer orang lain. Jika ada hanya menjelaskan tentang apa yang dimaksud dengan "Alat telekomunikasi", dimana pada pasal 1 UU RI No.36 tahun 1999 yang dimaksud dengan alat telekomunikasi adalah :

> "Alat telekomunikasi adalah setiap alat perlengkapan yang digunakan dalam bertelekomunikasi;"

Komputer dalam hal ini bisa dikategorikan sebagai "alat telekomunikasi", namun tidak ada pasal atau kalimat yang mengatur tentang pelarangan atau penghukuman dalam menggunakan alat telekomunikasi yang bukan miliknya secara tanpa hak akses/tidak sah/manipulasi !

Lalu bagaimana dengan 3 pertanyaan terakhir :
- Perbuatan tanpa hak akses ke jasa telekomunikasi
- Perbuatan tidak sah akses ke jasa telekomunikasi
- Perbuatan manipulasi akses ke jasa telekomunikasi

Mengacu pada arti kalimat "jasa telekomunikasi" pada Pasal 1 UU RI NO. 36 tahun 1999 yang menjelaskan tentang arti kata kalimat jasa telekomunikasi sebagai berikut :
> "Jasa telekomunikasi adalah layanan telekomunikasi untuk memenuhi kebutuhan bertelekomunikasi dengan menggunakan jaringan telekomunikasi"

Jelas sekali tersirat arti kalimat "jasa telekomunikasi" pada pasal 1 UU RI No. 36 tahun 1999 ini. Dari UU RI No.36 tahun 1999 ini pun tidak terdapat definisi yang jelas tentang istilah apa yang dimaksud dengan istilah "layanan telekomunikasi". Bila kita bisa definisikan maka "layanan telekomunikasi" adalah sarana dan prasarana untuk terlayan-nya telekomunikasi, biasanya berupa software.
Pertanyaannya apakah benar Dani menggunakan "layanan telekomunikasi" secara tanpa akses/tidak sah/manipulasi ?
Kalau dia menggunakan software Microsoft bisa jadi, jika komputer tempat dimana dia mengakses menggunakan software yang tidak license dari Microsoft, akan tetapi apa yang dituntut jauh melenceng, dituntut bukan karena membobol situs KPU tapi dituntut karena menggunakan software yang tidak license.

Jadi secara jelas pasal 22 UU RI No. 36 tahun 1999 tidak bisa digunakan sebagai pasal untuk menjerat dani !

Bagaimana tentang tuntutan pasal 38 UU RI No. 36 tahun 1999
Pasal 38 berbunyi:

Setiap orang dilarang melakukan perbuatan yang dapat menimbulkan gangguan fisik dan elektromagnetik terhadap penyelenggaraan telekomunikasi.

Apakah Dani melakukan perbuatan fisik, merusak secara langsung dan kelihatan secara fisik menghancurkan, atau membom terhadap sarana dan prasarana telekomunikasi ?
Apakah Dani melakukan perbuatan yang menimbulkan gangguan elektromagnetik ? Apakah Dani menimbulkan gangguan pada transmisi gelombang data ?
Sekali lagi Dani hanya mengakses komputer dimana data-data tentang hasil Pemilu terekam dan tidak merusak secara fisik ataupun elektromagnetik terhadap sarana dan prasarana telekomunikasi, dan jelas pada UU RI No. 36 tahun 1999 tidak ada kalimat atau pasal yang dapat menjerat Dani karena mengakses dan merubah isi data komputer yang menyimpan data-data KPU secara tidak sah atau tanpa hak. Penekannya adalah "merubah" jika mengakses setiap orang berhak untuk mengakses data dari Komputer yang menyimpan data-data KPU yang secara nyata memang bertujuan akan diumumkan oleh KPU kepada masyarakat luas

Jadi jelas Dani tidak bisa dituntut oleh pasal 38 UU RI NO.36 tahun 1999 karena isi atau makna pasal 38 UU RI No. 36 tahun 1999 ini terlalu lemah!

Bagimana dengan tuntutan pasal 50 UU RI NO.36 tahun 1999
Pasal 50 berbunyi:

Barang siapa yang melanggar ketentuan sebagaimana dimaksud dalam Pasal 22, dipidana dengan pidana penjara paling lama 6 (enam) tahun dan atau denda paling banyak Rp600.000.000,00 (enam ratus juta rupiah).

Jelas sekali Dani tidak melanggar pasal 22 UU RI No. 36 tahun 1999, oleh karena itu secara nyata Dani tidak dapat dipidana dengan pidana penjara paling lama 6 (enam) tahun dan atau denda paling banyak Rp600.000.000,00 (enam ratus juta rupiah).